\documentclass[letterpaper,english,reprint, aps, nofootinbib]{revtex4-1}
\usepackage[T1]{fontenc}
\usepackage[latin9]{inputenc}
\usepackage{xcolor}
\usepackage{verbatim}
\usepackage{amsmath}
\usepackage{amssymb}

\makeatletter

\providecommand{\tabularnewline}{\\}

\usepackage{graphicx}
\usepackage{xcolor}
\usepackage[bookmarks=false,linkcolor=blue,urlcolor=blue,colorlinks,citecolor=blue]{hyperref}
\usepackage{natbib}

\makeatother

\usepackage{babel}
\usepackage{soul}
\setstcolor{red}

\begin{document}

\preprint{}

\title{Quantum Anomalous Parity Hall Effect in Magnetically Disordered Topological Insulator Films}

\author{Arbel Haim$^{1,2}$, Roni Ilan$^{3}$, and Jason Alicea$^{1,2}$}

\affiliation{%
	\begin{tabular}{c}
		$^{1}$Department of Physics and Institute for Quantum Information
		and Matter,\tabularnewline
		California Institute of Technology, Pasadena, California 91125, USA \tabularnewline
		$^{2}$Walter Burke Institute of Theoretical Physics, California Institute
		of Technology, Pasadena, California 91125, USA\tabularnewline
		$^{3}$Raymond and Beverly Sackler School of Physics and Astronomy,
		Tel-Aviv University, Tel-Aviv 69978, Israel\tabularnewline
\end{tabular}}

\date{\today}
\begin{abstract}
	In magnetically doped thin-film topological insulators, aligning the
	magnetic moments generates a quantum anomalous Hall phase supporting a single
	chiral edge state. We show
	that as the system de-magnetizes, disorder from randomly oriented magnetic moments can produce
	a `quantum anomalous parity Hall' phase with \emph{helical} edge
	modes protected by a unitary reflection symmetry. We further show that introducing superconductivity, combined with selective
	breaking of reflection symmetry by a gate, allows for creation and
	manipulation of Majorana zero modes via purely electrical means and at zero applied magnetic field.
\end{abstract}
\maketitle

{\bf \emph{Introduction}.}~Thin films of magnetically doped topological insulators (TIs) provide an experimental realization of the quantum anomalous Hall (QAH) effect~\citep{Yu2010quantized,Jiang2012quantum,Wang2013Quantum,Wang2013anomalous,Wang2014universal,Wang2015quantum,Chang2013experimental,Checkelsky2014trajectory,Kou2014scale,Bestwick2015Precise,Chang2015high,Kandala2015giant,Chang2016observation,Mani2018are,Mani2018role},
wherein a quantized Hall response emerges in the absence of
an external magnetic field. 
For a `pure' TI thin film (without magnetic dopants), the top and bottom surfaces host Dirac cones~\citep{Fu2007topological,Zhang2009topological,Hsieh2008topological,Xia2009observation,Chen2009experimental,Moore2007topological} that can gap out via hybridization through the narrow bulk---naturally yielding a trivial insulator. When the Zeeman energy from polarized magnetic moments overwhelms the inter-surface hybridization, the TI film instead enters
a QAH phase that exhibits a nontrivial Chern number together with a single chiral edge state that underlies conductance quantization. Studies of the magnetic structure~\citep{Lachman2015visualization,Lachman2017observation,Grauer2015coincidence}
suggest that the magnetic dopants form weakly interacting, nanometer-scale islands and interact via easy-axis ferromagnetic coupling within each island.
In typical experiments, these islands---which generically exhibit different coercive fields---are polarized by an external
magnetic field, though ferromagnetic interactions allow the sample to remain magnetized even as the field is eliminated.  

\begin{figure}
	\begin{centering}
		\begin{tabular}{lr}
			\includegraphics[clip=true,trim=0mm -15mm 0mm 0mm,width=4.2cm]{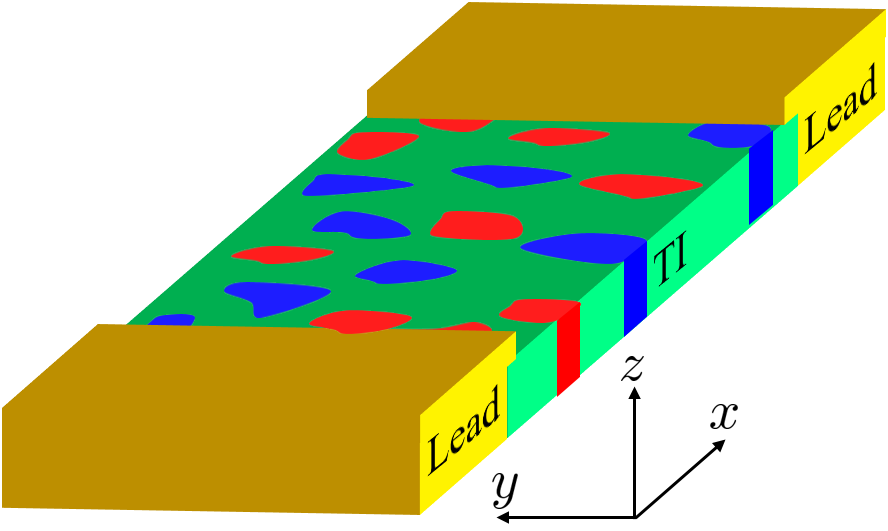}
			\llap{\hskip 0mm \parbox[c]{8.5cm}{\vspace{0cm}(a)}}
			&
			\hskip -3mm
			\includegraphics[clip=true,trim=3.5cm 8.7cm 4.5cm 8.3cm,width=4.5cm]{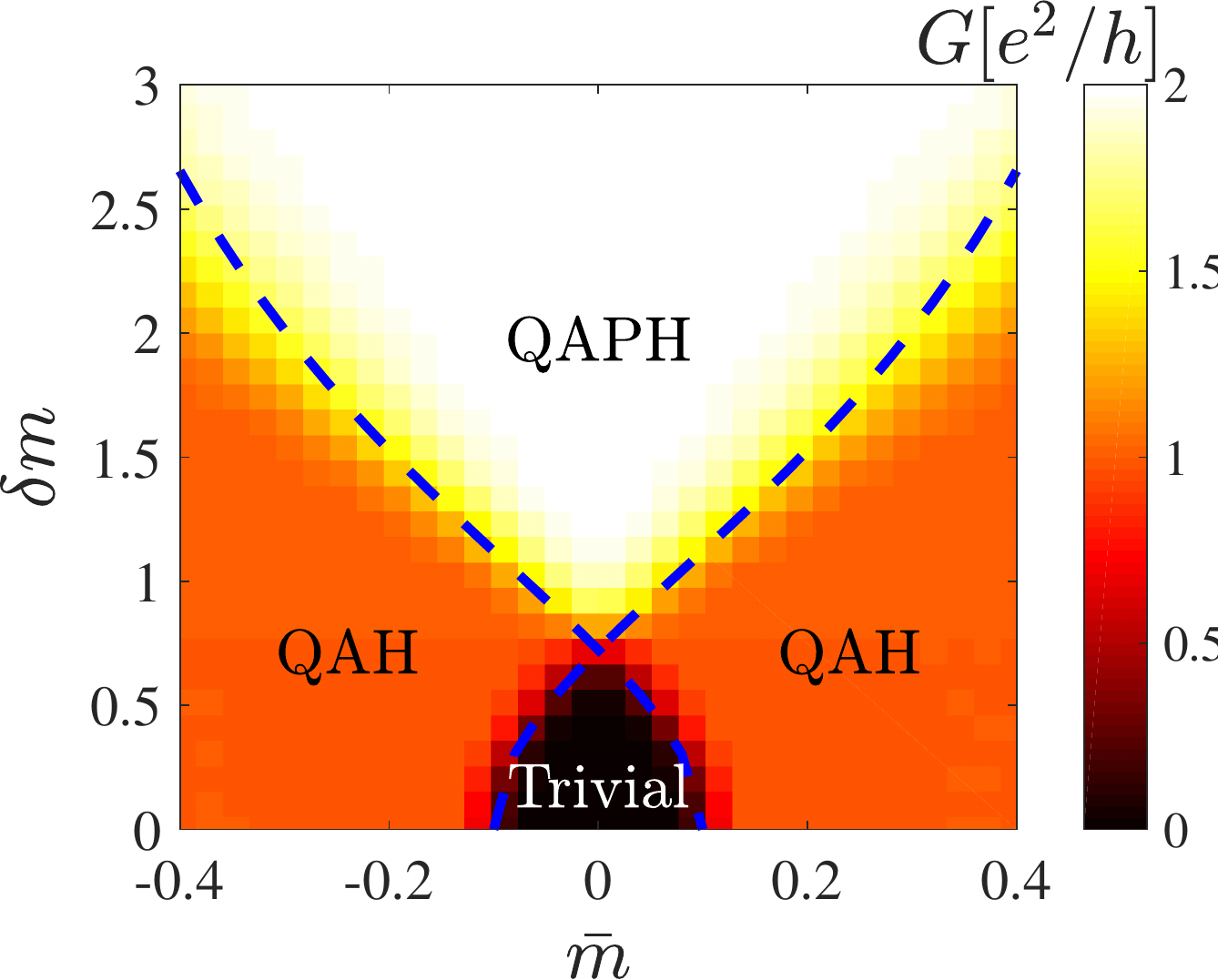}
			\llap{\parbox[c]{8.55cm}{\vspace{0cm}(b)}}
		\end{tabular}
	\end{centering}
	
	\caption{(a) Magnetically doped TI thin film probed by metallic leads.  Red and blue regions denote islands in which the magnetization points along $+\hat{z}$ and $-\hat{z}$, respectively.  (b) Phase diagram of the TI film
		as a function of the average magnetization, $\bar{m}=\left\langle m(x,y)\right\rangle$,
		and the magnetic-disorder strength, quantified by $\delta m=\sqrt{\left\langle m^{2}(x,y)\right\rangle -\bar{m}^{2}}$; color represents the two-terminal conductance extracted from Eq.~(\ref{eq:H_latt}) using parameters $v=2$, $t_{2}=1$,
		$t_{0}=0.1$, $\xi=2t_{2}/v$, $E_{\rm F}=0$, and $L_x=L_y=150$~\cite{FilmParamsUnits}. Sufficiently strong magnetic disorder drives the system into a `quantum anomalous parity Hall' (QAPH) phase
		with helical edge modes that are protected by $z\rightarrow-z$ reflection symmetry and yield quantized $2e^2/h$ conductance.  At weak magnetic disorder, either a QAH phase or trivial phase emerges depending on the average magnetization.  Blue dashed lines indicate approximate analytical phase boundaries obtained within the self-consistent Born approximation; see Eq.~(\ref{eq:phase_bound}).
		\label{fig:setup_and_phase_diag}}
\end{figure}

In this Letter we examine the de-magnetization process for the TI film, focusing on the regime in which the net magnetization vanishes. While one might expect that a trivial insulator supplants the QAH phase here, we show that a more interesting scenario can quite naturally emerge.  
In particular, a TI film with zero net magnetization experiences strong \emph{magnetic disorder} and features locally polarized magnetic domains that cancel only on average.  
We show that
this magnetic disorder can drive the system into a quantum-spin-Hall-like phase (first described in Ref.~\onlinecite{Hattori2015edge} in a different context) supporting
\emph{helical} edge modes that can be detected via standard transport measurements.  
Unlike the canonical quantum-spin-Hall state that is protected by time-reversal symmetry~\cite{Kane2005quantum,Bernevig2006quantum,Konig2007quantum}, these modes are protected
by a unitary reflection symmetry that interchanges the top and bottom surfaces.  The local magnetization in a thin TI film
is not expected to vary appreciably along the perpendicular direction [see Fig.~\hyperref[fig:setup_and_phase_diag]{\ref{fig:setup_and_phase_diag}(a)}];
the magnetically disordered film can
then at least approximately preserve this reflection symmetry under appropriate gating conditions. 

The physics we uncover can be viewed as a disorder-driven, zero-field counterpart to the very recently reported `quantum parity Hall effect' in trilayer graphene, where edge channels protected by mirror symmetry arise~\cite{Stepanov2019quantum} (see also Ref.~\onlinecite{Yafis2018Counterpropagating}). We therefore refer to our helical phase as a `quantum anomalous parity Hall' (QAPH) state. As an appealing application, we argue that the helical edge channels in a magnetically disordered TI film provide an ideal venue for pursuing Majorana zero modes (MZMs)~\cite{Alicea2010Majorana,Beenakker2013,Aguado2017majorana,Lutchyn2018majorana}. In a usual quantum-spin-Hall state, Majorana zero modes bind to domain walls separating regions of the edge gapped by proximity-induced superconductivity and by time-reversal symmetry breaking~\cite{Fu2009josephson}. Our QAPH phase requires only breaking of reflection symmetry---thereby eschewing applied magnetic fields altogether and enabling dynamical manipulation of Majorana zero modes using purely electrical means.

{\bf \emph{Model}.}~We consider a magnetically doped thin TI film \citep{Zhang2010crossover,Li2010chern,Shan2010effective,Hattori2015edge,Zhang2013electric}
described at low energies by 
\begin{equation}
\mathcal{H}=v(k_{x}\sigma_{x}+k_{y}\sigma_{y})\tau_{z}+t(k)\tau_{x}+m(\boldsymbol{r})\sigma_{z}.
\label{eq:H}
\end{equation}
Here $\boldsymbol{r} = (x,y)$ is a coordinate along the film, $k_{x,y} = -i \partial_{x,y}$ is the momentum, $\sigma_{x,y,z}$ are Pauli matrices acting in spin space,
and $\tau_{x,y,z}$ are Pauli matrices in the basis of states
belonging to the top and bottom surfaces. The first term encodes the Dirac spectrum for each surface ($v$ is the velocity).  The second hybridizes the two surfaces with tunneling matrix element $t(k)=t_{0}+t_{2}k^2$.  In the last term,
$m(\boldsymbol{r})$ is a Zeeman field induced by easy-axis magnetic dopants; note that the Zeeman field depends on $\boldsymbol{r}$ but is identical for the top and bottom surfaces.
Upon disorder averaging we assume
\begin{equation}
\left\langle m(\boldsymbol{r})\right\rangle =\bar{m}\,,\,\left\langle m(\boldsymbol{r})m(\boldsymbol{r}')\right\rangle -\bar{m}^{2}=\delta m^2 K(\boldsymbol{r}-\boldsymbol{r}'),
\end{equation}
where $K(\boldsymbol{r}-\boldsymbol{r}')$ decays with correlation length $\xi$ and is normalized so that $K(0) = 1$.  With this normalization the disorder strength is set by $\delta m$.  

Equation~(\ref{eq:H}) commutes with $\mathcal{M}=\tau_{x}\sigma_{z}$,
which implements a reflection about the $(x,y)$ plane. In the basis
that diagonalizes $\mathcal{M}$, the Hamiltonian therefore acquires
a block diagonal form, ${\mathcal{H}}=h_{+}\oplus h_{-}$, with
\begin{equation}
h_{\pm}=-v(k_{x}\sigma_{x}+k_{y}\sigma_{y})+[m(\boldsymbol{r})\pm t(k)]\sigma_{z}.
\label{eq:h_plus_minus}
\end{equation}
Each block describes a single Dirac cone with a disordered mass term. As an illuminating primer, let us examine the clean
limit where $m(\boldsymbol{r})=\bar{m}$. We assume a regularization of Eq.~\eqref{eq:h_plus_minus} such that the Chern number for block $h_{\pm}$ in this case is given by~\citep{Bernevig2006quantum,Qi2011topological}
\begin{equation}
C_{\pm}=\left[{\rm sgn}(\bar{m}\pm t_{0})\mp{\rm sgn}(t_{2})\right]/2.
\label{eq:ChernNumbers}
\end{equation}
For $|\bar{m}|>|t_{0}|$, only one of the blocks has zero Chern number, 
and the overall Chern number is $C={\rm sgn}(\bar{m})$.
This regime corresponds to the QAH phase that hosts a single chiral edge mode. For $|\bar{m}|<\left|t_{0}\right|$, the total Chern number necessarily vanishes.  When $t_0 t_2 > 0$ a trivial phase with $C_{\pm} = 0$ arises.  However, if $t_{0}t_{2}<0$ then the two blocks have
nonzero and opposite Chern number: $C_{\pm}= \pm {\rm sgn}(t_{0})$.
Here the system realizes a pristine QAPH phase supporting helical edge modes, with a right-mover coming from one block and a left-mover from the other. These edge
modes are protected from gapping only when reflection symmetry
is maintained.  Henceforth we will assume $t_0 t_2 >0$---which precludes the QAPH phase in the clean limit.  Below we show that introducing magnetic disorder through a spatially varying $m(\boldsymbol{r})$ nevertheless stabilizes the QAPH phase in a mechanism akin to that of the `topological
Anderson insulator'~\citep{Li2009topological,Jiang2009numerical,Groth2009theory,Prodan2011Prodan,Yamakage2011disorder,Xing2011topological}.

{\bf \emph{Analysis of magnetic disorder}.}~We now restore spatially non-uniform $m(\boldsymbol{r})$ in Eq.~\eqref{eq:H}. The phase boundaries separating the QAH, trivial, and QAPH states highlighted above can be analytically estimated using the self-consistent Born approximation, wherein disorder effects are captured by a self-energy term $\Sigma_{\pm}(\omega, \boldsymbol{k})$ associated with block $h_\pm$. The self-energy follows from the self-consistent equation
\begin{equation}
\begin{split}\Sigma_{\pm}(\omega,\boldsymbol{k})=\delta{m}^2\int&\frac{{\rm d}^{2}q}{(2\pi)^{2}}\tilde{K}(\boldsymbol{q}) \sigma_{z}\left[\omega-\bar{h}_\pm(\boldsymbol{k}-\boldsymbol{q})\right.\\
&\left. -\Sigma_{\pm}(\omega,\boldsymbol{k}-\boldsymbol{q})\right]^{-1}\sigma_{z}.
\end{split}\label{eq:SCBA}
\end{equation}
Here $\tilde{K}(\boldsymbol{q})$ is the Fourier transform of
$K(\boldsymbol{r})$, and $\bar h_\pm$ is defined as $h_\pm$ evaluated with $m(\boldsymbol{r})\rightarrow \bar m$.  Hereafter we set $\omega = 0$, which allows us to extract the Chern numbers for the disordered system from an effective Hamiltonian $h_{\pm}^{{\rm eff}}(\boldsymbol{k})=\bar h_{\pm}(\boldsymbol{k})+\Sigma_{\pm}(\omega = 0,\boldsymbol{k})$.

To facilitate analytical progress, we choose the function describing
disorder correlations to be $K(\boldsymbol{r})=2(\xi/r)J_{1}(r/\xi)$,
where $J_1$ is a Bessel function.  The Fourier transform then takes a particularly simple form: $\tilde{K}(\boldsymbol{q})=4\pi\xi^2\Theta(1-\xi q)$ with
$\Theta(x)$ the Heaviside step function. The low-momentum expansion of the self-energy takes the form~\cite{SM}
\begin{equation}\label{eq:Sigma_expansion}
\Sigma_{\pm}(0,\boldsymbol{k})=\Sigma_\pm^{(1)} (k_x \sigma_x + k_y \sigma_y) + 
(\Sigma_\pm^{(0)} +\Sigma_\pm^{(2)}\boldsymbol{k}^2)\sigma_z,
\end{equation} 
which allows us to obtain corrections to $v$, $t_2$, and $\bar{m}\pm t_0$. We are after the critical disorder strength, $\delta m_{\rm c}^\pm$, at which $h^{\rm eff}_\pm$ changes Chern number. This transition occurs when $\Sigma_\pm^{(0)}=-(\bar{m}\pm t_0)$. Using Eq.~(\ref{eq:Sigma_expansion}) and expanding the right-hand side of Eq.~(\ref{eq:SCBA}) to $\mathcal{O}(k^2)$ yields~\cite{SM}
\begin{equation}
\delta{m}_{\pm}^{{\rm c}}=\left[\frac{t_{2,\pm}^\prime\left(t_{0}\pm\bar{m}\right)}{\xi^2\ln\left\{ 1+\left[t_{2,\pm}^\prime/\left(v_\pm^\prime\xi\right)\right]^{2}\right\} }\right]^{1/2},\label{eq:phase_bound}
\end{equation}
where $v^\prime_\pm,t_{2,\pm}^\prime$ are the disorder-renormalized $v$ and $t_2$ parameters~\cite{SM}. Note that $v^\prime_\pm \approx v$ and $t_{2,\pm}^\prime \approx t_2$ when either $(\bar{m}\pm t_{0})t_{2}/v^{2}$ or  $v\xi/t_{2}$ are sufficiently small.

Dashed lines in Fig.~{\hyperref[fig:setup_and_phase_diag]{\ref{fig:setup_and_phase_diag}(b)} sketch the corresponding phase boundaries (see caption for parameters).  
Most interestingly, when $\delta{m}>\delta{m}_{+}^{{\rm c}},\delta{m}_{-}^{{\rm c}}$, the two blocks have nontrivial and opposite Chern numbers, and the system realizes the QAPH phase as advertised.  
If instead $\delta{m}<\delta{m}_{+}^{{\rm c}},\delta{m}_{-}^{{\rm c}}$ a trivial insulator emerges; otherwise the QAH state appears.

The physical picture underlying Eq.~(\ref{eq:phase_bound}) can be understood from the limit of long disorder-correlation length,
$\xi\gg v/\delta m, |t_2|/v$. Suppose first that $t_0=\bar{m}=0$ and $t_2=0^+$.
In this limit, the $h_+$ block in Eq.~\eqref{eq:h_plus_minus} describes domains of characteristic size $\xi$ with either magnetization $\delta m$ (yielding trivial Chern number $C_+ = 0$) or $-\delta m$ (yielding $C_+ = -1$); a chiral edge state propagates along each domain wall, reflecting the change in Chern number.  
Since the typical sizes for trivial and topological domains are equal here, the $h_+$ block overall is critical, in accordance with the percolation picture of the Chalker-Coddington network model~\citep{Chalker1988percolation}.

We can then examine the effect of a finite $t_2>0$ on the position of the boundary mode between two domains, described by the Schr\"odinger equation $iv\sigma_{x}\partial_{x}|\phi(x)\rangle+\left[\delta m\cdot {\rm sgn}(x)-t_{2}\partial_{x}^{2}\right]\sigma_{z}|\phi(x)\rangle=0$ where $x=0$ is taken as the boundary (see the Supplemental Material for details~\citep{SM}). The position of the edge mode with respect to the boundary can be quantified by the difference between the decay lengths towards either side of the boundary, $\Delta x = \lambda_+-\lambda_-$, where $\lambda_\pm = \lim_{x\to\pm\infty}\left| x/\ln[|\phi(x)|^2] \right|$. One obtains that to first order in $t_{2}\delta m/v^{2}$, the edge state shifts into the trivial domain by a distance $\Delta x=t_{2}/v$ [see Fig.~\hyperref[fig:long_xi_limit]{\ref{fig:long_xi_limit}(a)}], thereby enlarging the topological region and pushing the block into the topological phase.  Alternatively, introducing small but finite average magnetization and inter-surface tunneling, $\bar{m},t_0 \neq 0$, instead shifts the edge state by $\Delta x=-v(t_0+\bar{m})/\delta{m}^2$. The phase transition therefore occurs when these shifts cancel, corresponding to $\delta{m}^c_+= v[(t_0+\bar{m})/t_{2}]^{1/2}$, which indeed agrees with Eq.~\eqref{eq:phase_bound} in the limit of $\xi\gg t_2/v$ and $(\bar{m}+t_{0})t_{2}/v^{2}\ll1$.  Similarly, for the $h_-$ block [Fig.~\hyperref[fig:long_xi_limit]{\ref{fig:long_xi_limit}(b)}] one finds a transition at $\delta m_-^c = v[(t_0-\bar{m})/t_{2}]^{1/2}$, again in agreement with Eq.~\eqref{eq:phase_bound}.  

\begin{figure}
	\begin{centering}
		\begin{tabular}{lr}
			\hskip 2mm
			\includegraphics[clip=true,trim=0mm 0mm 0mm 0mm,height=3.55cm]{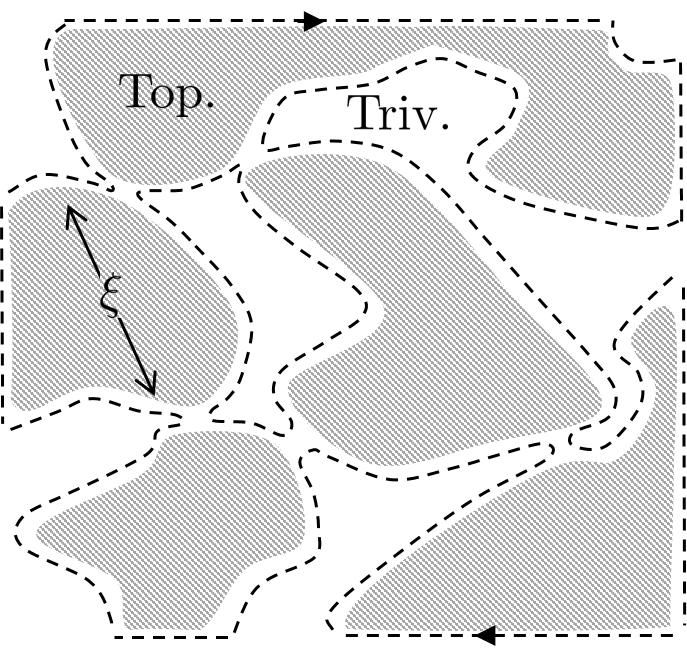}
			\llap{\parbox[c]{7.7cm}{\vspace{2mm}(a)}}
			&
			\hskip 4mm
			\includegraphics[clip=true,trim=0mm 0mm 0mm 0mm,height=3.45cm]{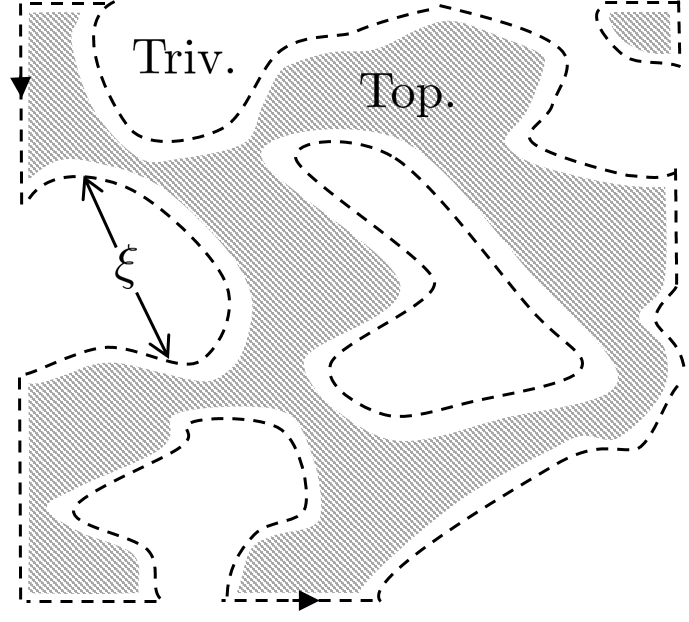}
			\llap{\parbox[c]{8.3cm}{\vspace{2mm}(b)}}
		\end{tabular}
	\end{centering}
	\vspace{-1mm}
	
	\caption{\label{fig:long_xi_limit} Physical picture for the quantum anomalous parity Hall phase in the long disorder-correlation-length limit. (a) At $t_0 = \bar m = 0$ and $t_2 = 0^+$, the sector described by $h_+$ realizes randomly oriented trivial (white) and topological (shaded) domains occupying equal areas, yielding critical behavior.  Finite $t_2$ shifts the edge states at the domain walls by a distance $r_\perp$ (see dashed line), thus enlarging the topological domains and pushing the $h_+$ block into the topological phase. (b) A similar picture holds for the $h_-$ block, but with trivial and topological domains switched and the edge-state chirality reversed.}
\end{figure}

The above analytic results can be corroborated numerically. To this end we discretize Eq.~(\ref{eq:H})
on an $L_{x}\times L_{y}$ square lattice, resulting in 
\begin{equation}
\begin{split}H & =\sum_{n_{x}=1}^{L_{x}}\sum_{n_{y}=1}^{L_{y}}\Big\{c_{\boldsymbol{n}}^{\dagger}\left(t_{0}'\tau_x + m_{\boldsymbol{n}}\sigma_z\right)c_{\boldsymbol{n}}\\
+& \sum_{\boldsymbol{d}=\hat{x},\hat{y}} \left[c_{\boldsymbol{n}}^{\dagger} \left(i\frac{v}{2}\boldsymbol{\sigma}\cdot\boldsymbol{d}\tau_z-t_{2}\tau_{x}\right)c_{\boldsymbol{n}+\boldsymbol{d}} + h.c.\right]\Big\},
\end{split}
\label{eq:H_latt}
\end{equation}
where $c_{\boldsymbol{n},\rho,s}^{\dagger}$ creates an electron with
spin $s$ on site $\boldsymbol{n}$ of surface $\rho$ (we suppress $\rho, s$ indices above);
lengths are measured in units of the lattice constant; and $t_{0}'=t_{0}+4t_{2}$.
The Zeeman field $m_{\boldsymbol{n}}$ is now taken to be normally distributed, with
average $\left\langle m_{\boldsymbol{n}}\right\rangle =\bar{m}$ and
correlations $\left\langle m_{\boldsymbol{n}}m_{\boldsymbol{n}'}\right\rangle -\bar m^{2}=\delta m^{2}\exp\left\{ -\left[\left(\boldsymbol{n}-\boldsymbol{n}'\right)/2\xi\right]^{2}\right\} $. 
We connect the system to two leads, as depicted
in Fig.~\hyperref[fig:setup_and_phase_diag]{\ref{fig:setup_and_phase_diag}(a)}, and compute the scattering
matrix for incoming and outgoing electrons~\citep{SM} using a recursive procedure that involves
gradually increasing the system's length in the $x$ direction~\cite{Lee1981Anderson}. The two-terminal
conductance obtained from the Landauer-B\"uttikker formalism
is $G=\left(e^{2}/h\right){\rm Tr}(t^{\dagger}t),$ where $t$
is the transmission matrix between the leads for electrons at the Fermi energy, $E_{\rm F}$. 

The color map in Fig.~\hyperref[fig:setup_and_phase_diag]{\ref{fig:setup_and_phase_diag}(b)} displays the conductance $G$
versus $\bar{m}$ and $\delta m$.  Each edge mode contributes $e^2/h$ to the conductance.  The trivial, QAH, and QAPH phases are thus readily diagnosed by quantized conductances $G = 0$, $e^2/h$, and $2e^2/h$, respectively.  
In the clean limit ($\delta m=0$)
one obtains the familiar scenario where the system passes from the trivial to the QAH phase
phase when $\left|\bar{m}\right|>t_{0}$.  Magnetic disorder instead drives the system into the QAPH state, as found analytically; note the good agreement between the analytical and numerical phase boundaries, despite the different disorder correlations used.

{\bf \emph{Reflection symmetry breaking}.}~To study the effects of breaking the reflection symmetry $\mathcal{M}=\tau_{x}\sigma_{z}$ that protects the QAPH edge states, we include an electric potential near the sample boundary that is opposite for the top and bottom surfaces; experimentally such a term can be controllably generated via asymmetric gating of the TI film.  We specifically perturb Eq.~\eqref{eq:H} [or its lattice counterpart, Eq.~\eqref{eq:H_latt}] with $\mathcal{H}'=V_{{\rm A}}(\textbf{r})\tau_{z}$, where 
$V_{\rm A}(\boldsymbol{r})=V_{\rm A}^0$ within a distance $W_{\rm edge}$ from the edges and $V_{\rm A}(\boldsymbol{r})= 0$ otherwise. Figure~\hyperref[fig:symmetry_breaking]{\ref{fig:symmetry_breaking}(a)} presents the two-terminal
conductance versus $V_{{\rm A}}$ for different linear system sizes $L_{x} = L_y \equiv L$, assuming system parameters corresponding to the QAPH phase. For $V_{{\rm A}}^0=0$, the conductance
reaches the quantized value of $G=2e^{2}/h$ as expected, independent
of system size. Increasing $V_{{\rm A}}^0$ generates backscattering among the helical edge states and thus reduces the conductance.  The system-size dependence is further explored in Fig.~\hyperref[fig:symmetry_breaking]{\ref{fig:symmetry_breaking}(b)}, which plots the conductance versus $L$ for three values of $V_{{\rm A}}^0$.
For a given $V_{{\rm A}}^0\neq0$, the probability of edge-mode backscattering increases with system size, 
thereby decreasing the conductance---albeit rather slowly.  Interestingly, since the counter-propagating
modes are not related by symmetry, they generally do not overlap in
space (see also~\citep{SM}). This property can suppress backscattering by
reflection-symmetry-breaking terms such as $V_{{\rm A}}^0$.

\begin{figure}
	\begin{centering}
		\begin{tabular}{lr}
			\hskip -1mm
			\includegraphics[clip=true,trim=36mm 80mm 43mm 89mm,height=3.5cm]{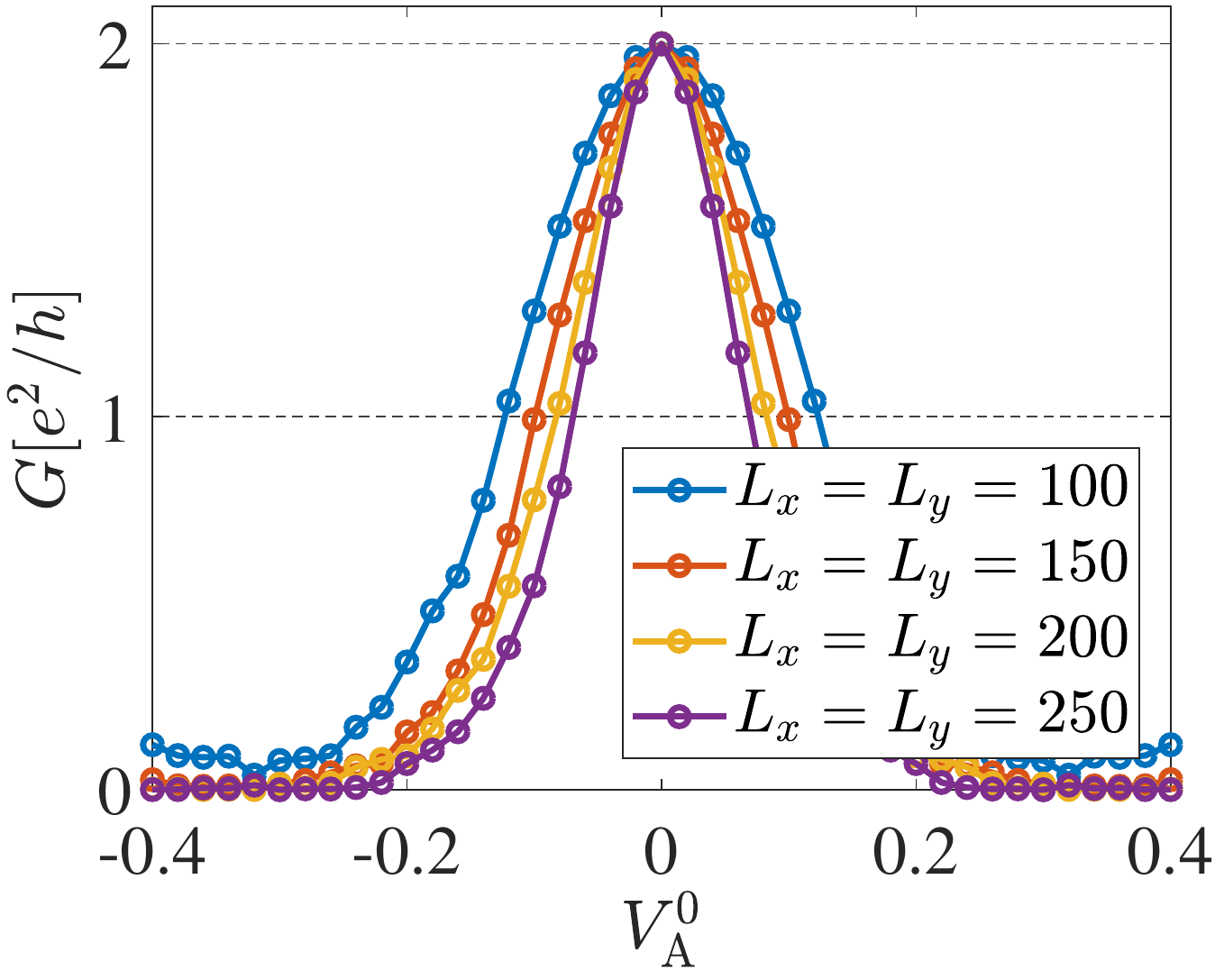}
			\llap{\parbox[c]{8.5cm}{\vspace{-1mm}(a)}}
			&
			\hskip -2mm
			\includegraphics[clip=true,trim=36mm 80mm 43mm 89mm,height=3.5cm]{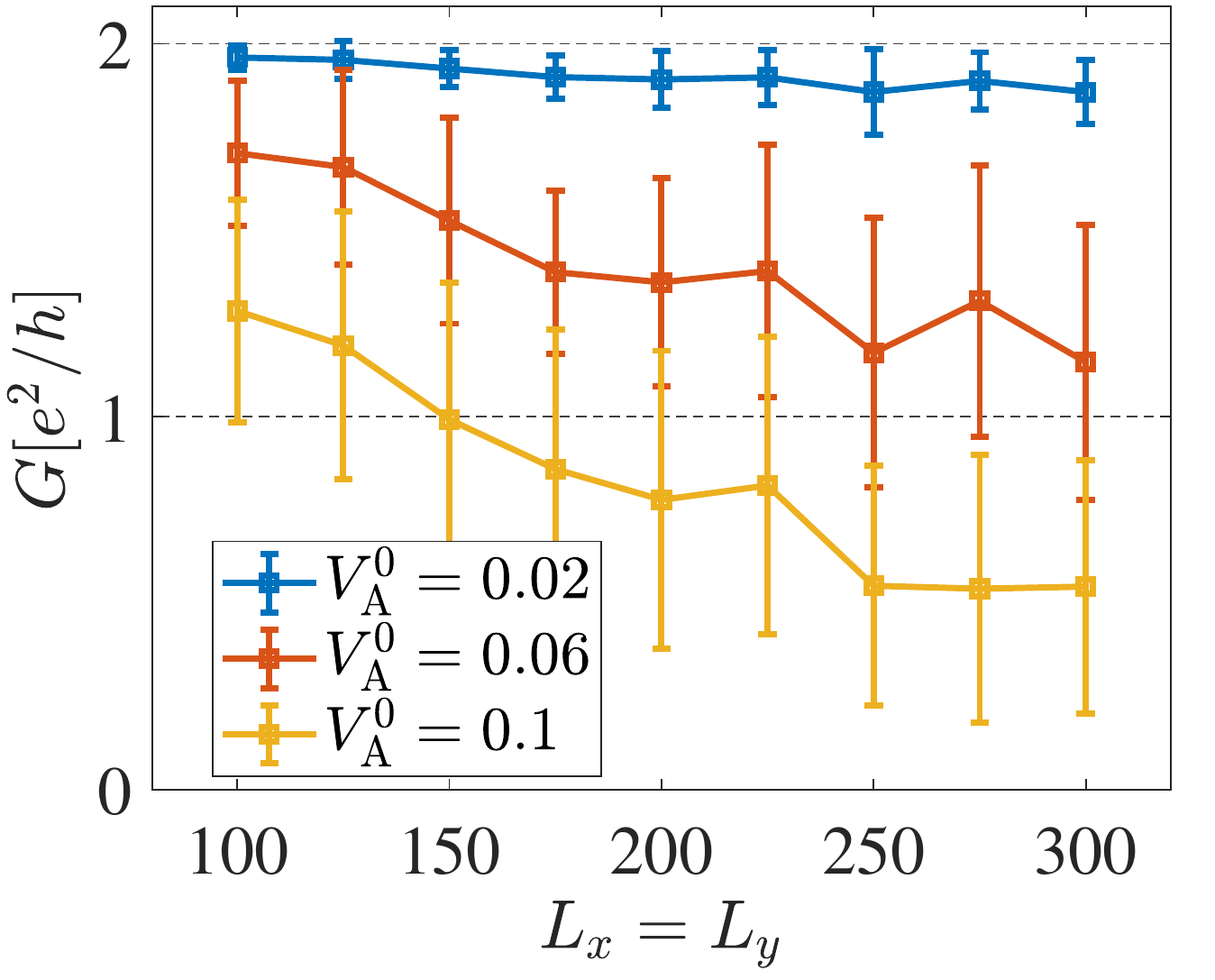}
			\llap{\parbox[c]{8.5cm}{\vspace{-1mm}(b)}}
		\end{tabular}
	\end{centering}
	\vspace{-2mm}
	
	\caption{(a) Conductance $G$
		versus reflection-symmetry-breaking potential $V_{\rm A}^0$ applied near the edge,
		for $\bar m=0$, $\delta m=2$, $\xi=\sqrt{2}t_{2}/v$, $E_{\rm F}=0.05$, $W_{\rm edge}=10$ and for different
		system sizes, $L_{x},L_{y}$.  Other parameters are the same as in
		Fig.~\hyperref[fig:setup_and_phase_diag]{\ref{fig:setup_and_phase_diag}(b)}. Each data point is a result of averaging over 50 disorder realizations. Nonzero $V_{\rm A}^0$
		allows backscattering between the helical edge modes, and therefore
		suppresses the conductance below the quantized $2e^{2}/h$ value. (b) Dependence of $G$ on system
		size for different $V_{\rm A}^0$ values; error bars represent the standard deviation.
		\label{fig:symmetry_breaking}}
\end{figure}

{\bf \emph{Superconductivity and Majorana modes}.}~The helical edge states in the QAPH phase 
serve as a natural platform for realizing MZMs upon coupling the edge to a conventional superconductor \citep{Fu2009josephson}.
In this setup a MZM localizes to the boundary between a section of the edge gapped by
superconductivity and a section gapped due to a reflection-symmetry-breaking potential $V_{{\rm A}}$ (see above).
The latter gapped regions can be accessed by making $V_{\rm A}$ arbitrarily large \emph{without} deleteriously impacting the parent superconductor---contrary to applied magnetic fields which alternative approaches typically require~\cite{Fu2009josephson,Oreg2010helical,Lutchyn2010majorana,Pientka2013Topological}.  Furthermore, 
locally controlling $V_{\rm A}$ through gates enables all-electrical manipulation of MZMs.

To demonstrate the realization of MZMs, we simulate
superconductivity in the setup from Fig.~\hyperref[fig:Adding_SC_AR]{\ref{fig:Adding_SC_AR}(a)} by
adding a pairing term, $\Delta_{\boldsymbol{n}}c_{\boldsymbol{n}\uparrow}^{\dagger}c_{\boldsymbol{n}\downarrow}^{\dagger}+{\rm h.\,c.}$,
to Eq.~(\ref{eq:H_latt}). Proximitizing the superconductor generally also induces an asymmetry in the chemical potential, which is simulated by a term $V_{{\rm A},\boldsymbol{n}}\boldsymbol{c_n}^\dagger\tau_z\boldsymbol{c_n}$. The potentials $\Delta_{\boldsymbol{n}}$ and $V_{{\rm A},\boldsymbol{n}}$ assume the values $\Delta_0$ and $V_{\rm A}^0$ beneath the superconductor but otherwise vanish.  
We then recalculate the scattering matrix, which now includes a block $r_{\rm he}$ describing Andreev reflection~\citep{SM}.  Figure~\hyperref[fig:Adding_SC_AR]{\ref{fig:Adding_SC_AR}(b)} presents the total Andreev reflection, $R_{{\rm he}}={\rm Tr}[r_{{\rm he}}^{\dagger}r_{{\rm he}}^{\phantom{\dagger}}]$,
for incident electrons at zero energy, for different values of $V_{{\rm A},\boldsymbol{n}}$ (see caption for parameters).
For $\left|\bar{m}\right|\lesssim0.8$, the system forms a QAPH 
phase whose helical edge states are gapped by superconductivity, and accordingly Andreev reflection occurs with near-unit probability.  The perfect Andreev reflection at zero energy signals the emergence
of a MZM at the boundary of the superconducting section of the edge
\citep{Law2009majorana,Fidkowski2012universal}.  For $\bar{m}\gtrsim0.8$,
the QAH phase appears, accompanied by a precipitous suppression of $R_{\rm he}$ due to the inability of a chiral mode to be reflected (either through normal or Andreev processes).

\begin{figure}
	\begin{centering}
		\begin{tabular}{lr}
			\includegraphics[clip=true,trim=2mm -15mm 0mm 0mm,width=4.25cm]{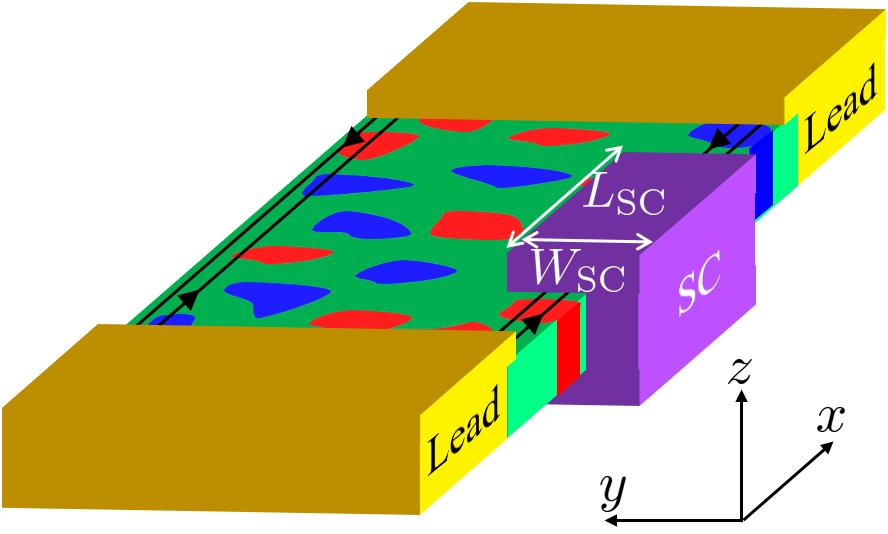}
			\llap{\parbox[c]{8.2cm}{\vspace{-6.5cm}(a)}}
			&
			\hskip -3mm
			\includegraphics[clip=true,trim=38mm 82mm 43mm 89mm,height=3.6cm]{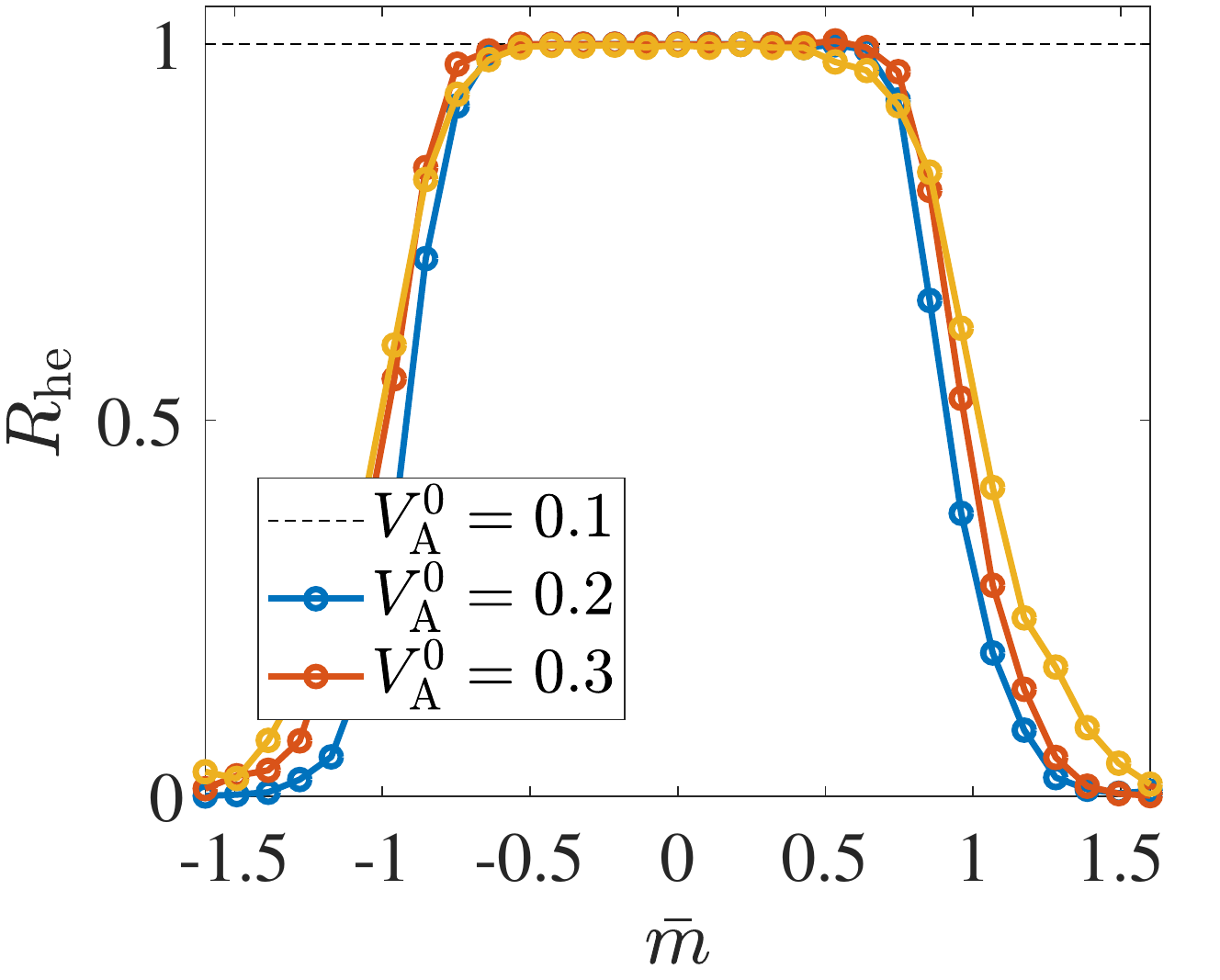}
			\llap{\parbox[c]{8.7cm}{\vspace{-6.5cm}(b)}}
		\end{tabular}
	\end{centering}
	\vspace{-3mm}
	
	\caption{(a) QAPH phase whose helical edge states are gapped by proximity-coupling to a conventional superconductor.  
		Majorana zero modes (MZMs) form at the edges of such pairing-gapped regions.   (b) Andreev reflection $R_{\rm he}$ for electrons incident at zero energy from the bottom lead as a function
		of the average Zeeman field, $\bar{m}$.  Data correspond to 
		magnetic disorder of strength $\delta m=2.5$, correlation length $\xi=\sqrt{2}t_{2}/v$, superconductor dimensions $W_{\rm SC}=50$, $L_{\rm SC}=80$, Fermi energy $E_{\rm F}=0$, and pairing potential $\Delta_0=0.5$, and are averaged over 50 disorder realizations; other 
		parameters are the same as in Fig.~\ref{fig:symmetry_breaking}.  
		When the system forms the QAPH state ($|\bar{m}|\lesssim0.8$),
		the MZM that emerges between the lead and the superconductor induces
		perfect Andreev reflection.\label{fig:Adding_SC_AR} }
\end{figure}

{\bf \emph{Discussion}.}~We have shown that experimentally motivated disorder originating from randomly oriented magnetic islands in TI thin films can stabilize a QAPH state.  This phase harbors helical edge states that are protected by a reflection symmetry that interchanges the top and bottom surfaces, and can be straightforwardly detected: In a two-terminal measurement, the QAPH state is characterized by a quantized $2e^2/h$ conductance that is \emph{enhanced} compared to that of the proximate QAH phases; recall Fig.~\hyperref[fig:setup_and_phase_diag]{\ref{fig:setup_and_phase_diag}(b)}. In a Hall-bar measurement, the QAPH should appear as a $\sigma_{xy}=0$ plateau~\cite{Feng2015observation}, together with $\sigma_{xx}=2e^2/h$. 

Breaking reflection symmetry suppresses the conductance below $2e^2/h$, though edge conduction should
still be observable in a finite system (see Fig.~\ref{fig:symmetry_breaking}).
Tuning in and out of the reflection-symmetric regime, while keeping
the chemical potential fixed, can be achieved by employing both bottom and top gates.  
We argued that the ability to electrically control the helical edge modes in this manner renders the proximitized QAPH system an ideal venue for exploring MZMs. Furthermore, having local control over the breaking of reflection symmetry (e.g. using gates) can allow for binding fractional charges at domain walls between regions where the reflection-symmetry-breaking term switches sign, analogous to the bound states discussed in Ref.~\onlinecite{Qi2008fractional}.

\begin{figure}
	\begin{centering}
		\begin{tabular}{lr}
			\includegraphics[clip=true,trim=0mm 0mm 0mm 0mm,height=2.6cm]{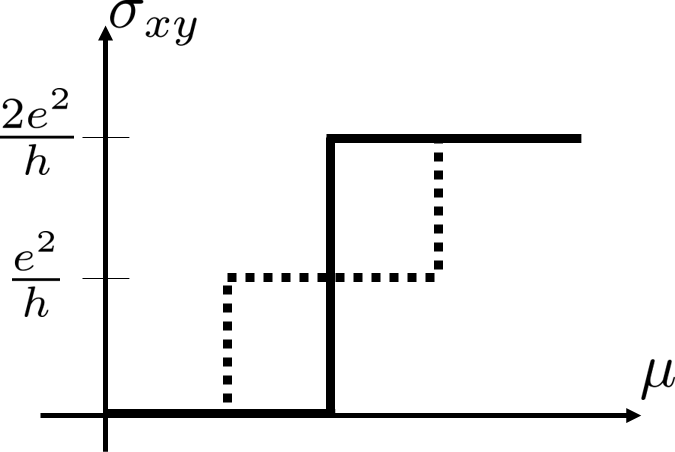}
			\llap{\parbox[c]{7.7cm}{\vspace{-45mm}(a)}}
			&
			\hskip 2mm
			\includegraphics[clip=true,trim=0mm 0mm 0mm 0mm,height=2.6cm]{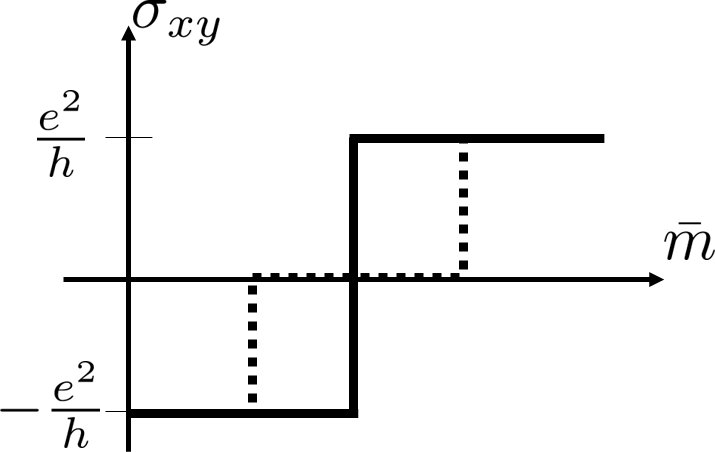}
			\llap{\parbox[c]{7.7cm}{\vspace{-45mm}(b)}}
		\end{tabular}
	\end{centering}
	\vspace{-2mm}
	
	\caption{(a) Adjusting the chemical potential $\mu$ to populate a spin-degenerate Landau level changes $\sigma_{xy}$ from 0 to $2e^2/h$ in the clean limit (solid line).  Spin-non-conserving disorder splits the plateau transition and opens an intervening phase with $\sigma_{xy} = e^2/h$ (dashed line) \cite{Lee1994unified}.  (b) In
		a magnetic TI film with no inter-surface tunneling, the picture is `shifted': Flipping the magnetization $\bar m$ takes the system between QAH phases with $\sigma_{xy} = -e^2/h$ and $+e^2/h$ in the clean limit (solid line).  Magnetic disorder can open an intervening phase with $\sigma_{xy} = 0$ (dashed line) corresponding to a QAPH state protected by reflection symmetry.\label{fig:spin_QH_analogy}}
\end{figure}

We close by providing a complementary perspective on our findings.  In a clean integer quantum Hall system, populating a spin-degenerate Landau level changes the Hall conductivity $\sigma_{xy}$ from 0 to $2e^2/h$.  Including disorder that breaks spin conservation generically splits this plateau transition and opens an intervening quantum Hall phase with $\sigma_{xy} = e^2/h$ \cite{Lee1994unified}; see Fig.~\hyperref[fig:spin_QH_analogy]{\ref{fig:spin_QH_analogy}(a)}.  In a clean magnetic TI film with no inter-surface tunneling, reversing the magnetization similarly changes $\sigma_{xy}$ from $-e^2/h$ to $+e^2/h$.  By analogy with the plateau transition, it is natural to expect that disorder can open up an intervening phase with $\sigma_{xy} = 0$~\cite{Wang2014universal} as sketched in Fig.~\hyperref[fig:spin_QH_analogy]{\ref{fig:spin_QH_analogy}(b)}.  Interestingly, we have shown that this disorder-induced phase need not be trivial---even though the Hall conductivity vanishes---when reflection symmetry is present.  This viewpoint may prove useful for discovering other disorder-induced symmetry-protected phases, e.g., in bands with higher Chern numbers.

{\bf \emph{Acknowledgments}.}~We thank A. Thomson for her insightful comments. We acknowledge support from the Army Research Office under Grant Award
W911NF-17-1-0323 (J.A.); the NSF through grants DMR-1723367 (J.A.);
the Caltech Institute for Quantum Information and Matter, an NSF Physics
Frontiers Center with support of the Gordon and Betty Moore Foundation
through Grant GBMF1250 (J.A.); and the Walter Burke Institute
for Theoretical Physics at Caltech (J.A. and A.H.).

\bibliographystyle{apsrev4-1}
\bibliography{References_ver_2}

\newpage
\begin{widetext}
	
\section*{Supplemental Material}

\section{Numerical simulation}

We begin by rewriting the lattice Hamiltonian, Eq.~(8) of the main
text, in the following form

\begin{equation}
H=\sum_{n_{x}=1}^{L_x}\left\{\vec{\psi}_{n_{x}}^{\dagger}h_{n_{x}}\vec{\psi}_{n_{x}}+\left[\vec{\psi}_{n_{x}}^{\dagger}V\vec{\psi}_{n_{x}+1}+{\rm h.c.}\right]\right\},
\end{equation}where $\{h_{n_{x}}\}_{n_{x}=1}^{L_{x}}$ and $V$ are $4L_{y}\times4L_{y}$
matrices, $\vec{\psi}_{n_{x}}^{\dagger}=(c_{(n_{x},1),{\rm b},\uparrow}^{\dagger},c_{(n_{x},1),{\rm b},\downarrow}^{\dagger},c_{(n_{x},1),{\rm t},\uparrow}^{\dagger},c_{(n_{x},1),{\rm t},\downarrow}^{\dagger},\dots,c_{(n_{x},L_{y}),{\rm b},\uparrow}^{\dagger},c_{(n_{x},L_{y}),{\rm b},\downarrow}^{\dagger}$, $c_{(n_{x},L_{y}),{\rm t},\uparrow}^{\dagger},c_{(n_{x},L_{y}),{\rm t},\downarrow}^{\dagger})$
is a $1\times4L_{y}$ vector of creation and annihilation operators, and `b' and `t' indicate bottom and top surfaces, respectively.

We place two normal-metal leads at $x=0$ and $x=L_{x}$. The reflection
matrix for electrons incident from the right can be calculated from the Green function, $G_{n_{x}}(\omega)$, using
~\cite{Fisher1981relation,Iida1990statistical}
\begin{equation}\label{eq:Weidenmuller}
r(\omega) = 1-2\pi i \rho_{\rm R}V^{\dagger}[G_{L_{x}}^{-1}(\omega)+i\pi \rho_{\rm R}VV^{\dagger}]^{-1}V,
\end{equation}where $\rho_{{\rm R}}$ is the density of states in the right lead, and
$G_{L_{x}}$ is the Green function matrix at the right-most sites
of the system, obtained through the recursive relation~\cite{Lee1981Anderson}
\begin{equation}\label{eq:Local_Green}
G_{n_{x}}(\omega)=[\omega-h_{n_{x}}-V^{\dagger}G_{n_{x}-1}V]^{-1}.
\end{equation}
For every $n_x=1,\dots,L_x$, the Green function $G_{n_{x}}(\omega)$ is a $4N_{y}\times4N_{y}$ matrix (indices running over `b'/`t', spin, and $n_{y}$), and $G_{0}=i\pi\rho_{{\rm L}}$, where $\rho_{{\rm L}}$ is the density of states in the left lead. The linear conductance is then given by
\begin{equation}
G\equiv\frac{{\rm d}I}{{\rm d}V}=\frac{e^{2}}{h}{\rm Tr}[t^{\dagger}t]=\frac{e^{2}}{h}{\rm Tr}[\boldsymbol{1}-r^{\dagger}r],
\end{equation}
where $t$ is the transmission matrix evaluated at the Fermi
energy $\omega=E_{{\rm F}}$.

When simulating the proximity-coupled superconductor, we amend the
Hamiltonian via $H\to H+\sum_{\boldsymbol{n},\rho}(\Delta_{\boldsymbol{n}} c_{\boldsymbol{n}\rho\uparrow}^{\dagger}c_{\boldsymbol{n}\rho\downarrow}^{\dagger}+{\rm H.c.}) + \sum_{\boldsymbol{n},s}V_{{\rm A},\boldsymbol{n}}(c_{\boldsymbol{n},{\rm b},s}^\dagger c_{\boldsymbol{n},{\rm b},s}-c_{\boldsymbol{n},{\rm t},s}^\dagger c_{\boldsymbol{n},{\rm t},s})$ and rewrite it in a Bogoliubov-de Gennes form,
\begin{equation}
H=\frac{1}{2}\sum_{n_{x}=1}^{L_x}\left\{\vec{\Psi}_{n_{x}}^{\dagger}\tilde{h}_{n_{x}}\vec{\Psi}_{n_{x}}+\left[\vec{\Psi}_{n_{x}}^{\dagger}\tilde{V}\vec{\Psi}_{n_{x}+1}+{\rm h.c.}\right]\right\},
\end{equation}
where $\vec{\psi}_{n_{x}}^{\dagger}=(\vec{\psi}_{n_{x}}^{\dagger},\vec{\psi}_{n_{x}})$,
and $\{\tilde{h}_{n_{x}}$\}, $\tilde{V}$ are $8N_{y}\times8N_{y}$
matrices. The reflection matrix, $\tilde{r}(\omega)$ is given by
Eqs.~\eqref{eq:Weidenmuller} and \eqref{eq:Local_Green}, with $\{h_{n_{x}}\}$
and $V$ replaced by $\{\tilde{h}_{n_{x}}\}$ and $\tilde{V}$ , respectively.
The diagonal blocks $r_{{\rm ee}}$ and $r_{{\rm hh}}$ describe
normal reflection, while the off-diagonal ones, $r_{{\rm he}}$ and
$r_{{\rm eh}},$ describe Andreev reflection. 

\subsection{Local density of states}

It was pointed out in the main text that the counter-propagating modes
in the QAPH phase are not related by symmetry, and therefore generally
do not overlap in space. This can be seen most conveniently in the
local density of states. In Fig.~\ref{fig:LDOS} we present the local
density of states, $\rho_{\pm}(x,y,\omega=0)$, corresponding to modes even
and odd under the reflection symmetry, calculated according to the method
described in Ref.~\citep{Potter2011Majorana}. The results are shown for a single disorder realization having $\bar{m}=0$ and $\delta m=2$. The rest of the system parameters are the same as in Fig.~1(b) of the main text. For these parameters, the system is in the QAPH phase. Both the $+$ and $-$ sectors host a mode confined roughly to the edge, though the positions clearly do not coincide.

\begin{figure}
	\begin{centering}
		\begin{centering}
			\begin{tabular}{lr}
				\includegraphics[clip=true,trim=30mm 80mm 30mm 80mm,height=6cm]{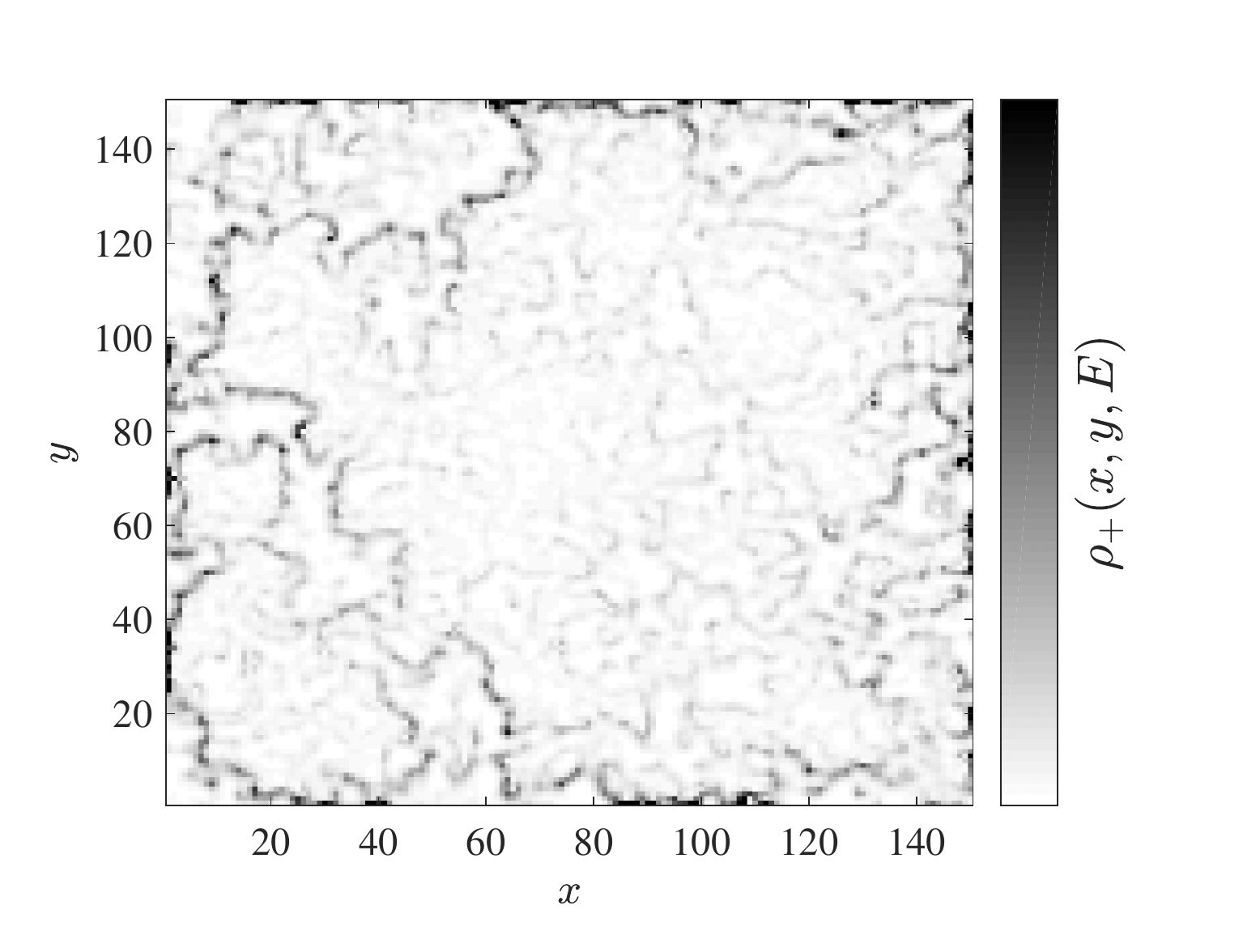}
				\llap{\parbox[c]{1.8cm}{\vspace{-10.5cm}High}}
				\llap{\parbox[c]{1.87cm}{\vspace{-1.7cm}Low}}
				\llap{\parbox[c]{15cm}{\vspace{-1cm}(a)}}
				&
				\hskip 0mm
				\includegraphics[clip=true,trim=30mm 80mm 30mm 80mm,height=6cm]{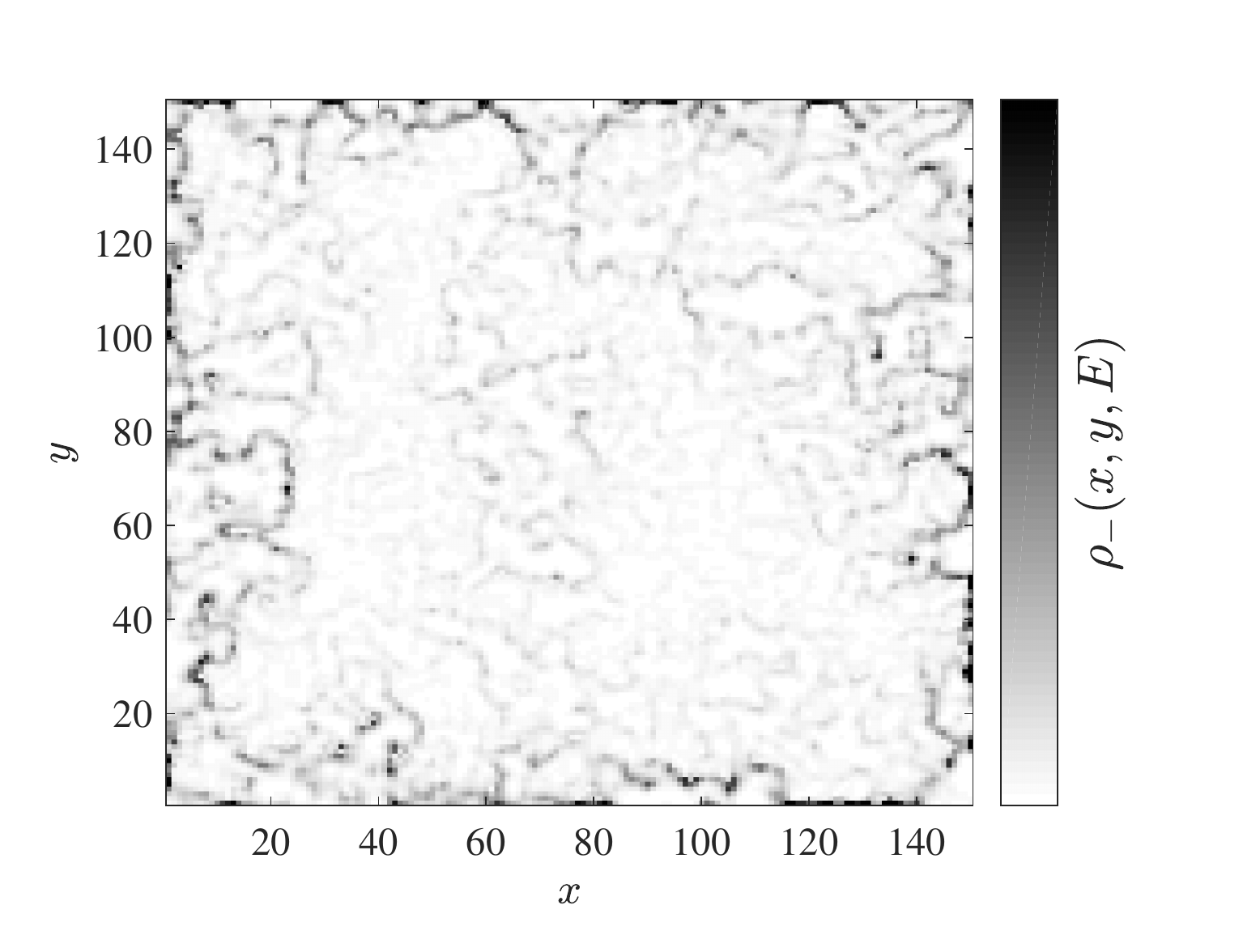}
				\llap{\parbox[c]{1.8cm}{\vspace{-10.5cm}High}}
				\llap{\parbox[c]{1.87cm}{\vspace{-1.7cm}Low}}
				\llap{\parbox[c]{15cm}{\vspace{-1cm}(b)}}
			\end{tabular}
		\end{centering}
		\par\end{centering}
	\caption{Local density of states in the QAPH phase, shown
		separately for modes that are (a) even and (b) odd under reflection symmetry $\mathcal{M}$.\label{fig:LDOS}}
\end{figure}

\section{Self-consistent Born approximation}

We start from Eq.~(5) of the main text and focus on the $h_{+}$
block (the treatment of the $h_{-}$ block follows along the same
lines). The self-energy can generally be written as
\begin{equation}
\Sigma_{+}(k,\omega)=\Sigma_{0}(k,\omega)+\Sigma_{x}(k,\omega)\sigma_{x}+\Sigma_{y}(k,\omega)\sigma_{y}+\Sigma_{z}(k,\omega)\sigma_{z}.
\end{equation}
Substituting the above expansion in Eq.~(5) of the main text results in four coupled
equations for $\Sigma_{0,x,y,z}(k,\omega)$. We are interested in
the self-energy at $\omega=0$ from which we define
an effective Hamiltonian,
\begin{equation}
h_{+}^{{\rm eff}}(\boldsymbol{k})=-vk_{x}\sigma_{x}-vk_{y}\sigma_{y}+\left[\bar{m}+t_{0}+t_{2}\boldsymbol{k}^{2}\right]\sigma_{z}+\Sigma_{+}(\boldsymbol{k},0).\label{eq:h_eff}
\end{equation}
For $\omega=0$, the self-consistent equations read
\begin{subequations}\label{eq:self_cons_comp}
	\begin{align}
	& \Sigma_{0}(\boldsymbol{k})=\frac{\delta m^{2}}{(2\pi)^{2}}\int\frac{{\rm d}^{2}q\tilde{K}(\boldsymbol{q})\Sigma_{0}(\boldsymbol{k}-\boldsymbol{q})}{\sum_{i=x,y}\left[v(k_{i}-q_{i})-\Sigma_{i}(\boldsymbol{k}-\boldsymbol{q})\right]^{2}+\left[\bar{m}+t_{0}+t_{2}(\boldsymbol{k}-\boldsymbol{q})^{2}+\Sigma_{z}(\boldsymbol{k}-\boldsymbol{q})\right]^{2}-\Sigma^2_{0}(\boldsymbol{k}-\boldsymbol{q})},\label{eq:self_cons_0}\\
	& \Sigma_{x,y}(\boldsymbol{k})=\frac{\delta m^{2}}{(2\pi)^{2}}\int\frac{{\rm d}^{2}q\tilde{K}(\boldsymbol{q})\left[-v(k_{x,y}-q_{x,y})+\Sigma_{x,y}(\boldsymbol{k}-\boldsymbol{q})\right]}{\sum_{i=x,y}\left[v(k_{i}-q_{i})-\Sigma_{i}(\boldsymbol{k}-\boldsymbol{q})\right]^{2}+\left[\bar{m}+t_{0}+t_{2}(\boldsymbol{k}-\boldsymbol{q})^{2}+\Sigma_{z}(\boldsymbol{k}-\boldsymbol{q})\right]^{2}-\Sigma^2_{0}(\boldsymbol{k}-\boldsymbol{q})},\label{eq:self_cons_xy}\\
	& \Sigma_{z}(\boldsymbol{k})=-\frac{\delta m^{2}}{(2\pi)^{2}}\int\frac{{\rm d}^{2}q\tilde{K}(\boldsymbol{q})\left[\bar{m}+t_{0}+t_{2}(\boldsymbol{k}-\boldsymbol{q})^{2}+\Sigma_{z}(\boldsymbol{k}-\boldsymbol{q})\right]}{\sum_{i=x,y}\left[v(k_{i}-q_{i})-\Sigma_{i}(\boldsymbol{k}-\boldsymbol{q})\right]^{2}+\left[\bar{m}+t_{0}+t_{2}(\boldsymbol{k}-\boldsymbol{q})^{2}+\Sigma_{z}(\boldsymbol{k}-\boldsymbol{q})\right]^{2}-\Sigma^2_{0}(\boldsymbol{k}-\boldsymbol{q})}.\label{eq:self_cons_z}
	\end{align}
\end{subequations}
We have omitted above the $\omega$ argument for brevity. 

First, we notice that Eq.~(\ref{eq:self_cons_0}) is satisfied by
$\Sigma_{0}(\boldsymbol{k})=0$. Second, in accordance with our long-wavelength
analysis of the system, we expand $\Sigma_{x,y,z}(\boldsymbol{k})$
to second order in $\boldsymbol{k}$,
\begin{equation}
\Sigma_{x,y}=\Sigma^{(1)}k_{x,y}\hspace{1em};\hspace{1em}\Sigma_{z}=\Sigma^{(0)}+\Sigma^{(2)}\boldsymbol{k}^{2}.
\end{equation}
The form of the above expansion derives from the symmetry properties
of the self-consistent equations, Eq.~(\ref{eq:self_cons_comp}),
whereby $(\Sigma_{x},\Sigma_{y})$ transform as $(k_{x},k_{y})$,
and $\Sigma_{z}$ transforms as a scalar under rotations and reflections
of $\boldsymbol{k}$. Plugging this expansion in Eq.~(\ref{eq:self_cons_comp}),
we have
\begin{align}
\Sigma^{(1)}\boldsymbol{k} & =-\frac{\delta m^{2}}{(2\pi)^{2}}\int\frac{{\rm d}^{2}q\tilde{K}(\boldsymbol{q})v'(\boldsymbol{k}-\boldsymbol{q})}{v'^{2}(\boldsymbol{k}-\boldsymbol{q})^{2}+\left[\bar{m}+t_{0}+\Sigma^{(0)}+t_{2}'(\boldsymbol{k}-\boldsymbol{q})^{2}\right]^{2}},\label{eq:self_cons_comp_exp}\\
\Sigma^{(0)}+\Sigma^{(2)}\boldsymbol{k}^{2} & =-\frac{\delta m^{2}}{(2\pi)^{2}}\int\frac{{\rm d}^{2}q\tilde{K}(\boldsymbol{q})\left[\bar{m}+t_{0}+t_{2}'(\boldsymbol{k}-\boldsymbol{q})^{2}\right]}{v'^{2}(\boldsymbol{k}-\boldsymbol{q})^{2}+\left[\bar{m}+t_{0}+\Sigma^{(0)}+t_{2}'(\boldsymbol{k}-\boldsymbol{q})^{2}\right]^{2}},\nonumber 
\end{align}
where $v'\equiv v-\Sigma^{(1)}$, and $t_{2}'=t_{2}+\Sigma^{(2)}$.

We are looking for the critical disorder strength, $\delta m_{\rm c}$, for which the $h_+$ block is at a transition between a trivial and a topological phase. This happens when the $\boldsymbol{k}=0$ coefficient of the $\sigma_z$ term in Eq.~(\ref{eq:h_eff}) vanishes, namely when $\Sigma_{z}(\boldsymbol{k}=0)=-\bar{m}-t_{0}$. We substitute this condition in Eq.~(\ref{eq:self_cons_comp_exp}), and solve the integrals to second order in $\boldsymbol{k}$, which results in the following equations
\begin{subequations}\label{eq:expan_ord_by_ord}	
	\begin{align}
	\Sigma^{(1)}\boldsymbol{k} & =-\frac{\delta m_{\rm c}^{2}\xi^{2}}{v'}\frac{1}{1+[t_{2}'/(v'\xi)]^{2}}\boldsymbol{k}
	+\mathcal{O}(\boldsymbol{k}^{3})\label{eq:expan_ord_by_ord_xy}
	\\
	-\bar{m}-t_{0}+\Sigma^{(2)}\boldsymbol{k}^{2} & =-\frac{\delta m_{{\rm c}}^{2}\xi^{2}}{t_{2}'}\ln\left\{ 1+\left[t_{2}'/(v'\xi)\right]^{2}\right\} +\frac{\delta m_{{\rm c}}^{2}t_{2}'^{3}/v'^{4}}{\left\{ 1+\left[t_{2}'/(v'\xi)\right]^{2}\right\} ^{2}}\boldsymbol{k}^{2}++\mathcal{O}(\boldsymbol{k}^{4}).\label{eq:expan_ord_by_ord_z}
	\end{align}
\end{subequations}
Finally, requiring that Eqs.~(\ref{eq:expan_ord_by_ord_xy},\ref{eq:expan_ord_by_ord_z})
are obeyed order by order results in the expression for the critical
disorder strength,
\begin{equation}
\delta m_{{\rm c}}^{2}=\frac{\left(\bar{m}+t_{0}\right)t_{2}'}{\xi^{2}\ln\left\{ 1+\left[t_{2}'/(v'\xi)\right]^{2}\right\} },
\end{equation}
and the following equations for $v',t_{2}'$,\begin{subequations}
	\begin{align}
	v'& = v+v'\frac{1}{\left[1+\left(\frac{t_{2}'}{v'\xi}\right)^{2}\right]\ln\left[1+\left(\frac{t_{2}'}{v'\xi}\right)^{2}\right]}\frac{\left(\bar{m}+t_{0}\right)t_{2}'}{v'^{2}},\label{eq:v}\\
	t_{2}' & =t_{2}+t_{2}'\frac{\left[t_{2}'/(v'\xi)\right]^{2}}{\left\{ 1+\left[t_{2}'/(v'\xi)\right]^{2}\right\} ^{2}\ln\left\{ 1+\left[t_{2}'/(v'\xi)\right]^{2}\right\} }\frac{\left(\bar{m}+t_{0}\right)t_{2}'}{v'^{2}}.\label{eq:t2}
	\end{align}
\end{subequations}
Note that Eqs.~(\ref{eq:v},\ref{eq:t2}) are solved by $v'\simeq v$, $t_{2}'\simeq t_{2}$ when either $(\bar{m}+ t_{0})t_{2}/v^{2}$ or  $v\xi/t_{2}$ are sufficiently small.

\section{The long-range-disorder limit}

As explained in the main text, in the large $\xi$ limit the results
can be understood by considering the effect of $t_{2}$ on the position
of the boundary mode between two domains. Consider, for example,
the sector $h_{+}$ and let us start from the
case $t_{0}=0$. We examine the boundary ($x=0$) between two domains
of opposite-sign mass, 
\begin{equation}
h_{+}=-v(k_{x}\sigma_{x}+k_{y}\sigma_{y})+[m(x)+t_{2}\boldsymbol{k}^{2}]\sigma_{z}.
\end{equation}
where $\bar{m}(x)=\bar{m}+\delta m\cdot{\rm sgn}(x),$ and $\delta m>|\bar{m}|$.
For simplicity of presentation we assume $t_{2}>0$,
which means that $x<0$ is topological, while $x>0$ is trivial.

We look for a mid gap solution to the Schr\"odinger equation,
\begin{equation}
iv\sigma_{x}\partial_{x}|\phi(x)\rangle+\left[m(x)-t_{2}\partial_{x}^{2}\right]\sigma_{z}|\phi(x)\rangle=0,
\end{equation}
where we have set $k_{y}=0$. The solution is given by
\begin{equation}
|\phi(x)\rangle=e^{\gamma_0 x}|\downarrow_{y}\rangle\begin{cases}
e^{-\gamma_{+}x}, & x\ge0\\
\cosh(\gamma_{-}x)-\frac{\gamma_{+}}{\gamma_{-}}\sinh(\gamma_{-}x), & x<0,
\end{cases}
\end{equation}
where $\gamma_0=\frac{v}{2t_{2}}$ and $\gamma_{\pm}=\frac{v}{2t_{2}}\sqrt{1+4t_{2}(\bar{m}\pm\delta m)/v^{2}}$.

The wave function decays away from the boundary at $x=0$, however, the decay lengths to the left and to the right are not equal. They are given by
\begin{equation}
\lambda_\pm = \lim_{x\to\pm\infty}\left| x/\ln[|\phi(x)|^2] \right|=\frac{1}{2|{\rm Re}(\gamma_0-\gamma_\pm)|}.
\end{equation}
When $\lambda_->\lambda_+$, the wave function is shifted towards the left, effectively decreasing the size of the topological region and overall driving $h_+$ to the trivial phase. When $\lambda_+>\lambda_-$, on the other hand, the wave function is shifted towards the right thereby increasing the size of the topological region and driving $h_+$ to the topological phase. This shift can be quantified by $\Delta x=\lambda_+-\lambda_-$, which in the limit of $\delta m t_2/v^2\ll1$ is given by
\begin{equation}
\Delta x = \frac{t_2}{v} - \frac{v\bar{m}}{\delta m^2 - \bar{m}^2}.
\end{equation} 
The transition between the trivial and topological phases of $h_+$ occurs when $\Delta x =0$, which to leading order in $\bar{m}t_2/v^2$ (recall $|\bar{m}|<\delta m$) is at the critical disorder strength
\begin{equation}
\delta m_{\rm c} = v \sqrt{\frac{\bar{m}}{t_2}}
\end{equation} 
Finally, the introduction of a finite $t_{0}$ is accounted for by
taking $\bar{m}\to\bar{m}+t_{0}$, while for the odd symmetry sector $h_{-}$ one takes $\bar{m}\to\bar{m}-t_{0}$ and $t_{2}\to-t_{2}$.
This reproduces Eq.~(7) of the main text in the limit of $\xi\gg t_{2}/v$ and $(t_{0}\pm\bar{m})t_{2}/v^{2}\ll1$.

\end{widetext}
	
\end{document}